\documentclass[aps,prb,superscriptaddress,twocolumn,showpacs,amsmath,amssymb,showkeys]{revtex4-1}
\usepackage{graphicx}
\usepackage{epstopdf}% Include figure files
\usepackage{xcolor}
\usepackage{amsmath}

\def\ns{\,\textrm{ns}}
\def\gcm{\,\textrm{g/cm}}
\def\tAA{\,\textrm{\AA}}
\def\eV{\,\textrm{eV}}
\def\K{\,\textrm{K}}

\begin{document}

\title{Machine Learning Assisted Modeling of Amorphous TiO$_2$-Doped GeO$_2$ \\ 
for Advanced LIGO Mirror Coatings}

\author{Jun Jiang}
\email[Email: ]{j.jiang@northeastern.edu}
\affiliation{Department of Physics, Northeastern University, Boston MA 02115, USA}
\affiliation{Department of Physics, University of Florida, Gainesville, Florida 32611, USA}

\author{Rui Zhang}
\affiliation{Department of Physics, Northeastern University, Boston MA 02115, USA}
\affiliation{Department of Physics, University of Florida, Gainesville, Florida 32611, USA}

\author{Kiran Prasai}
\affiliation{Department of Physics, Kennesaw State University, Marietta, GA 30060, United States of America}

\author{Riccardo Bassiri}
\affiliation{E. L. Ginzton Laboratory, Stanford University, Stanford CA 94305, USA}

\author{James N. Fry}
\affiliation{Department of Physics, Northeastern University, Boston MA 02115, USA}
\affiliation{Department of Physics, University of Florida, Gainesville, Florida 32611, USA}

\author{Martin M. Fejer}
\affiliation{E. L. Ginzton Laboratory, Stanford University, Stanford CA 94305, USA}

\author{Hai-Ping Cheng}
\email[Email: ]{ha.cheng@northeastern.edu}
\affiliation{Department of Physics, Northeastern University, Boston MA 02115, USA}
\affiliation{Department of Physics, University of Florida, Gainesville, Florida 32611, USA}

\begin{abstract}

The mechanical loss angle of amorphous TiO$_2$-doped GeO$_2$ can be lower than 10$^{-4}$, making it a candidate for Laser Interferometer Gravitational-wave Observatory (LIGO) mirror coatings. Amorphous oxides have complex atomic structures that are influenced by various factors, including doping concentration, preparation, and thermal history, resulting in different mass densities and physical properties. Modeling at atomistic level enables capturing these effects by generating atomic structure models according to experimental conditions. In order to obtain reliable and physical amorphous models at an affordable cost, we develop classical and machine-learning potentials (MLP) to speed up simulations. First-principles calculations are used to train and validate MLP as well as validating structure models. To better reproduce properties such as elastic modulus, radial distribution function (RDF) and the variations in mass density of doped amorphous oxides, density functional theory (DFT) calculations are used to optimize the final models. We find that the mass densities of amorphous systems are correlated with the total void volume. The experimental mass density matches the models with the most symmetric potential energy wells under volume change. The elastic response of the metal-oxygen network is also studied. The 27\% TiO$_2$ doped GeO$_2$ system shows the least number of large atom-atom distance changes, while for 44\% TiO$_2$ doped GeO$_2$, a majority of Ti-O distances are significantly changed. In response to strains, the metal-oxygen network at low mass densities prefers to adjust bond angles, while at high mass densities, the adjustment is mainly done by changing atom-atom distance.

\end{abstract}

\date{\today}
\maketitle

\section{Introduction}

Amorphous oxide mirror coatings are used in the Laser Interferometer Gravitational-wave Observatory (LIGO) due to their surface uniformity, low optical absorption and low mechanical loss. The current LIGO reflective mirror coatings have a multi-layer structure that consists of alternating high refractive index and low refractive index materials.\cite{kaiser2013optical,aasi2015advanced}  Amorphous TiO$_2$-doped Ta$_2$O$_5$ has proved to have both high reflective index, low optical absorption and low mechanical loss, hence has been used in LIGO high-reflective mirror coatings.\cite{harry2006titania,granata2020amorphous}

Thermal noise in the LIGO mirror coatings is one of the sensitivity-limiting noise sources\cite{saulson1990thermal,levin1998internal}. This noise can be characterized by internal mechanical loss. To further lower the mechanical loss of the mirror coatings, doped amorphous GeO$_2$ has been proposed as a candidate coating material.\cite{vajente2021low} Amorphous GeO$_2$ has very low mechanical loss at room temperature\cite{topp1996elastic}, suitable for LIGO mirror coatings, but the refractive index of pure amorphous GeO$_2$ is relatively low, and it cannot be used directly as the high refractive layer. Doping high reflective index oxides presents an option. Amorphous titanium dioxide is one of the best high reflective index materials, but can not be used standalone due to its low crystalline temperature\cite{wisniewski2022underestimated}. Doping amorphous TiO$_2$ into amorphous GeO$_2$ can increase the refractive index and achieve improved mirror coatings with lower mechanical loss compared to the previous best TiO$_2$-doped Ta$_2$O$_5$/silicon mirror coatings.\cite{vajente2021low} Amorphous oxides such as silica and titanium oxides are named a continuous random network (CRN), in which atoms are connected by covalent bonds (metal-oxygen bonds). TiO$_2$ doping will change the GeO$_2$ metal-oxygen network, elastic properties and thermal stability. Amorphous oxides, especially the doped amorphous oxides are high entropy materials, which means they can have stable strong local strains and stresses, which recently have been observed experimentally within high entropy alloys.\cite{moniri2023three} These internal strains accompanied with void spaces within doped amorphous oxides introduce additional complexity to the atomic structures, resulting in wide variation in mass densities. These internal strains are potentially related to the two-level system distribution and mechanical loss.\cite{bassiri2013correlations}

To study these complex doped amorphous oxides by atomistic modeling, a unified form of Morse-BKS potential has been developed\cite{van1990force,yu2009unified,trinastic2013unified}. These classical pair potentials combined with the reversed Monte Carlo method in Ta$_2$O$_5$, ZrO$_2$-doped Ta$_2$O$_5$\cite{prasai2019high,jiang2021analysis,mishkin2023hidden} have successfully captured the experimental radial distribution function (RDF). However, existing potentials have deficiencies. In ZrO$_2$-doped Ta$_2$O$_5$, the double peak feature of Ta-Ta pairs is missing using classical pair potentials\cite{jiang2023amorphous}, and in amorphous GeO$_2$, the neighboring Ge-Ge distance is shorter than that from the results of density functional theory (DFT). To improve the potentials, a three-body interaction term is added to the Morse-BKS classical pair potentials, giving better structures, also shown in previous study of $\alpha$-quartz SiO$_2$\cite{ghobadi2000crystal}. These problems can be overcome in first-principles calculations\cite{giacomazzi2006vibrational,marrocchelli2009construction,igram2018large}, but first-principles calculations with these large-sized amorphous atomic models are very time consuming and become unfeasible when the number of atoms goes over a thousand. Machine learning potentials are therefore good candidates, allowing one to speed up calculations while keeping the desired accuracy. For example, explicit multi-element spectral neighbor analysis potential (EME-SNAP) potentials\cite{cusentino2020explicit} have been developed specifically for amorphous ZrO$_2$-doped Ta$_2$O$_5$, which enable accurate molecular dynamic simulations using large models (18,000 atoms) where all atom-atom correlations match results from density functional theory (DFT) results.\cite{jiang2023amorphous}

In this work, we develop 1) classical potentials that add three-body interactions to the Morse-BKS potential for amorphous TiO$_2$-doped GeO$_2$ to enable fast MD simulations and 2) machine learning potentials to capture more structural features with acceptable speed. With these potentials, we can have atomic models with experimental mass density at local minimum, while classical potentials fail to do so. Atomic structures with various doping concentrations and mass densities are studied with accuracy close to that of DFT results.  Energetics, short- and medium-range orders, and elastic properties with different doping concentrations and mass densities are studied in detail.

\section{Computational Details}

\subsection{First-principle calculations}

First-principles calculations are performed using the Vienna Ab-initio Simulation Program (VASP) code\cite{kresse1993ab,kresse1996efficient,kresse1996efficiency} derived from self-consistent density functional theory (DFT)\cite{kohn1965self} using projector-augmented wave potentials (PAWs)\cite{blochl1994projector,kresse1999ultrasoft} in conjunction with the plane-wave expansion. The exchange and correlation functional is calculated using the parameter-free generalized gradient approximation (GGA) developed by Perdew, Burke, and Ernzerhof (PBE).\cite{perdew1996generalized} Only the gamma point is used in the calculations for amorphous models. We have tested energy cutoffs up to $ 800 \eV $ with no difference in the atomic structures and only slight shifts in potential energies. In this work, the energy cutoff is set to $ 520 \eV$. The maximum force during atomic structure relaxation is 0.02 eV/\AA.

\subsection{Classical Pair Potential}

Classical pair potentials that originate from physics understandings, such as nuclear repulsion, electronic bonding, and coulomb/dipole interactions, have recognizable analytic forms and are easy to calculate. They are robust and can be used in extreme conditions depending on the different potential parameters. Our first step is developing a new classical potential for amorphous GeO$_{2}$ that contains three body interactions based on the Morse-BKS form and is compatible with other Morse-BKS potentials. By simply mixing the same Morse-BKS form of different amorphous oxide potentials\cite{yu2009unified,trinastic2013unified}, the unified potentials allow us do atomic modeling with arbitrary combinations of metal elements in the amorphous oxides as long as the Morse-BKS potentials are available. The full equation of the Morse-BKS-3body potential is 
\begin{eqnarray}
&&U_{ij}(r_i,r_j)=\frac{q_i q_j}{r_{ij}}+a_{ij} e^{{-r_{ij}}/{\rho}}-\frac{c_{ij}}{r_{ij}^6}-D_{ij} \\ 
&&{}\qquad +D_{ij}[1-e^{-a_{ij}(r_{ij}-r_e)}]^2+
\sum_{jk} K(\theta_{ijk}-\theta_0)^2 , 
\nonumber
\end{eqnarray}
where $ \theta_{ijk} $ is the angle between bond $ij$ and bond $ik$, and the sum is over neighbors $j$, $k$.
In this work, we focus on the amorphous TiO$_2$-doped GeO$_2$ system. The existing Morse-BKS potentials for GeO$_2$ tend to overestimate the correlation of the Ge-Ge pairs\cite{peralta2008structural,hawlitzky2008comparative,marrocchelli2009construction}. The three-body interaction has been added to optimize the Morse-BKS potential for amorphous GeO$_2$, and improve the Ge-Ge distance. This Morse-BKS-3body potential for amorphous GeO$_2$ is compatible with the other Morse-BKS potentials for amorphous oxides, which can be used to generate initial models for doped amorphous oxides with correct pair correlations in terms of RDF. These especially guarantee the most important short-range order to be physically correct. Although the classical pair potentials are less accurate in terms of bond angle distributions, they are more stable for atomic structures that are far from local energy minimum structures or at high temperatures.

\subsection{Machine Learning Potential}

Machine learning (ML) is increasingly popular due to its ability to solve complex problems with proper training using existing knowledge. Machine learning potentials (MLP) for interatomic interactions are one ML application, using machine learned numerical models to predict potential energy surfaces for materials from atomic structures. MLP can be as accurate as first-principles calculations, while orders of magnitude faster \cite{zuo2020performance,jiang2023amorphous}. Due to their complicated energy landscape, modeling doped amorphous oxides is extremely difficult. Determining some properties requires a large simulation box to capture a fair sample of configurations. Hence, fast and accurate potentials (or force fields) are essential. Machine learning (ML) offers a powerful approach to generate inter-atomic potentials that are superior to the conventional classical functions used in simulations.ML Potentials enable simulations of large number of atoms accurately, especially for complex materials where current classical potentials are not available or fail to reproduce atomic structure features found from experiments. In this work, we generate different types of machine learning potentials for amorphous doped oxides (TiO$_2$-doped GeO$_2$): EME-SNAP \cite{cusentino2020explicit}, moment tensor machine-learning potential (MTP) \cite{shapeev2016moment,novikov2020mlip} and VASP on-the-fly MLP\cite{jinnouchi2019phase,jinnouchi2019fly,jinnouchi2020descriptors}. The on-the-fly MLP in VASP provide very accurate ML potentials for atomic models near the trained configurations, and they are much faster compared to DFT calculations but still significant slower compared to classical potentials, which limits their usage in large models and long time simulations. SNAP, and MTP are much faster compared to on-the-fly MLP in VASP. They are trained specifically for amorphous TiO$_2$-doped GeO$_2$ with doping levels ranging from 0--50\% to get both good atomic structure models and atomic densities compared to experiment. Simulations of rapid annealing and mechanical spectroscopy have also been performed successfully with these new machine learning potentials.

We compare the potential energies of each model based on DFT, SNAP and on-the-fly Gaussian Approximation Potential (GAP) machine learning potential (MLP): DFT using PBE-PAW; MLP machine learning potential on the fly generated by VASP from relaxed up to $2500\K$ MD trajectories; 
SNAP potential generated from relaxed, up to $2000\K$ MD trajectories, as well as relaxed models with single atom random shifted, volume changed from different dopings; 
and machine learning potential based on MTP MLIP with different cutoffs which show balancing the calculation errors and stability. MTP MLIP machine learning potentials are trained with the MLIP2 package\cite{novikov2020mlip}  running with the Large-scale Atomic/Molecular Massively Parallel Simulator (LAMMPS) software package\cite{plimpton1995fast}. SNAP machine learning potential are trained using FitSNAP\cite{rohskopf2023fitsnap}.

\begin{figure}
	\centering
	\includegraphics[width=0.40\textwidth]{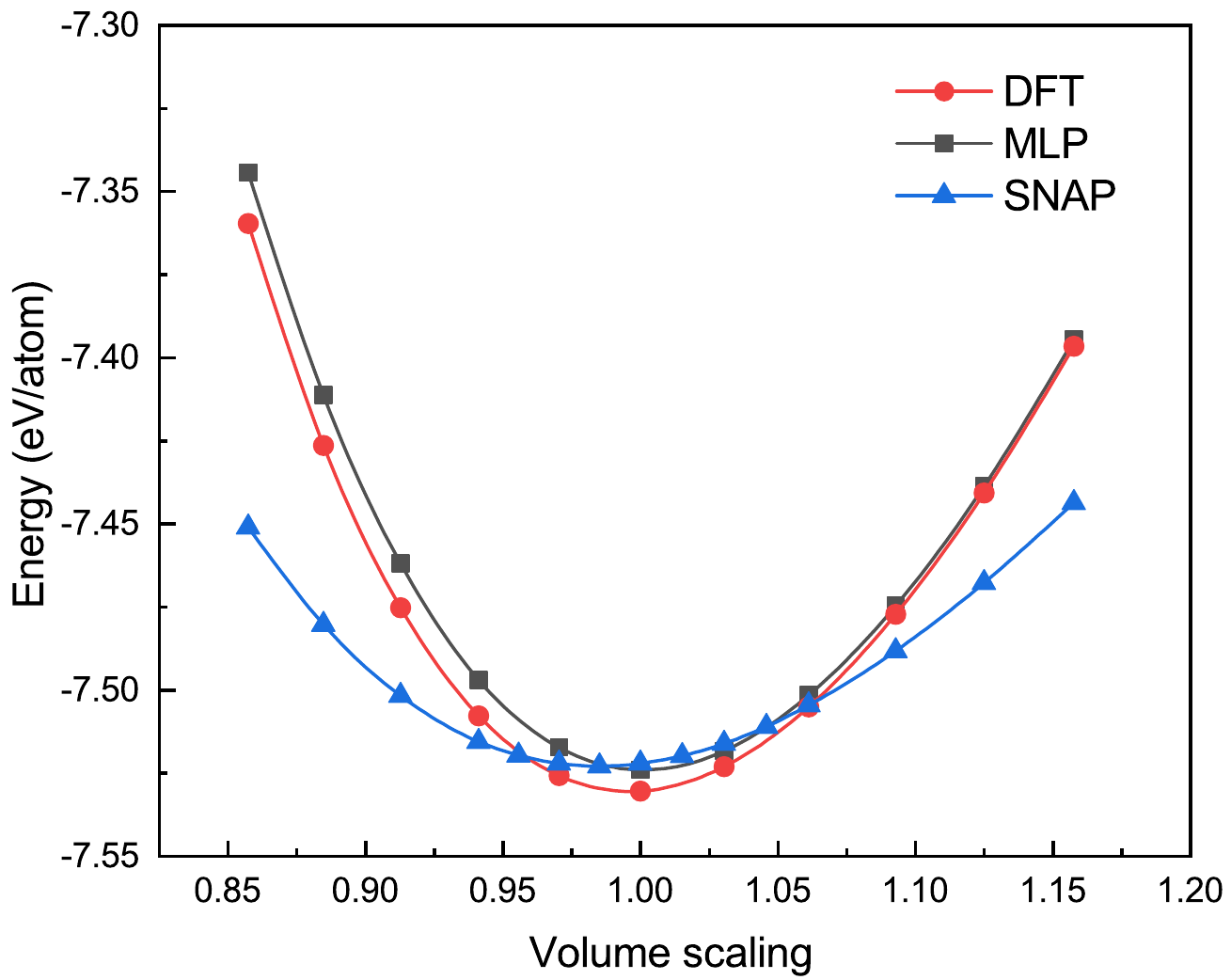}
	\caption{SNAP, VASP-MLP, and DFT energies of 48\% TiO$_2$-doped GeO$_2$ vs. volume change.}
	\label{energycompare}
\end{figure}

\subsection{Void Characterization}

Voids have been studied in the contexts of bio-materials and solids due to their importance in molecular geometry and materials properties such as defects and ion diffusivity\cite{levitt1992pocket,hendlich1997ligsite,till2010mcvol,beck2017nuclear,douglass2022vast}. In this work, we study voids and void distributions in amorphous oxides. We define a void space if one can potentially insert at least one rare gas atom, such as Ar. Experiments show that there are a few percent of Ar atoms embedded in the oxide materials during the ion beam sputtering process, and the Ar content is reduced during the thermal annealing process.\cite{paolone2022argon}

The volume of total voids depends on the mass density of the amorphous materials, and the void size distribution is a result of competition between surface energy versus energy cost due to strain in inter-atomic bonds. Void analysis can provide a way to get some insight into the metal-oxygen networks, such as uniformity and ring distributions.\cite{guttman1990ring,yang2021ring,shiga2023ring} These voids can also potentially absorb O atom and reorganize local atomic structures by lowering amorphous oxide potential energy.

\section{Results and discussion}

%The mechanical loss angle of amorphous GeO$_2$ can be lower than 10$^{-4}$ at room temperature\cite{topp1996elastic,vajente2021low}, which translates to low thermal noise in the LIGO mirror coatings. But, the relatively low refractive index (n $=$ 1.60 at 1064 nm) of amorphous GeO$_2$ requires thicker coatings to achieve the same reflectivity, which increases the total deposition time of the highly reflective (HR) coatings and also results in higher mechanical loss of the whole reflective coating. Since TiO$_2$ has high reflective index, doping with TiO$_2$ can significantly increase reflective index of amorphous GeO$_2$ without introducing excessive optical or mechanical losses. The good thermal stability of the amorphous oxides allows the deposited film to be annealed at higher temperatures without crystallization. Experimentally a-GeO$_2$ will crystallize annealed at $675$--$770 ^{\circ}$C\cite{yamaguchi1982crystallization} or at as-deposited condition within aqueous environments\cite{xiao2018phase}. TiO$_2$ doped amorphous GeO$_2$ can be thermally annealed to $600 ^{\circ}$C for 106 hours without crystallization\cite{vajente2021low,bhowmick2024unveiling}, allowing the thermal treatment to remove structure defects, optimize stoichiomety and reduce the mechanical/optical loss.
Classical potentials and machine learning potentials are used to assist the modeling and simulation processes based on first-principle calculations. The potential energies of TiO$_2$-doped GeO$_2$ with different doping concentrations and mass densities are calculated to explore the possible atomic structures and energy landscape. The voids, local strains and elastic responses are studied in detail in the following sections.  

\subsection{Accuracy of potentials vs computational cost}

The main reason to apply machine learning techniques for amorphous material modeling is due to the high calculation cost for first-principles calculations, while classical potentials are either not available or not accurate enough. Machine learning potentials are used to find numerical models which can balance calculation cost and accuracy. For complex materials, one needs to make compromises among calculation cost, accuracy, and versatility. A simple form of potential can be very accurate in modeling certain crystal, amorphous or liquid phases for systems such as noble gases. However, a simple classical potential often has much larger errors in phases for which properties are not targeted during construction of the potential.

On-the-fly machine learning potentials as implemented in VASP are based on the Gaussian aggregation method.\cite{jinnouchi2019phase,jinnouchi2019fly,jinnouchi2020descriptors} These on-the-fly potentials can be very accurate as long as calculations fall within the training terrain but they are relatively slow compared to others.  For doped amorphous oxides, they can only predict for certain doping concentrations within targeted temperatures. EME-SNAP, MTP and UF3 can give similar accuracy in energies and forces, but they also suffer from versatility and cannot predict different doping concentrations accurately with one trained MLP or achieve high temperature stability as good as classical pair potentials. We also find by allowing reduced accuracy, MTP can significantly improve stability at higher temperatures. In this work, we adopted a hierarchical approach to construct machine learning potentials to enable the generation and simulations of amorphous oxides accurately and efficiently. DFT calculations via VASP are accompanied with on-the-fly machine learning potentials to generate relative small models ($10^2$--$10^3$ atoms). These models are used to train less accurate but much faster EME-SNAP or MTP potentials, which are used to generate models with larger unit cells (more than $10^4$ atoms).

For the moment tensor machine-learning potential (MTP), we generate two different versions of potential; one is more accurate in predicting energies and forces near room temperature, while the other is more stable at high temperatures without completely breaking up the atomic structures. Chemical reactions can occur in the amorphous oxide materials; for example, it is well known that at higher temperatures, the oxides can lose or gain oxygen molecules.\cite{abernathy2021exploration} Experimentally, behavior also depends on how we deposit these oxides. The less accurate version of MLP is designed to forbid these possible chemical reactions, focusing on the amorphous phases we interested with a loss of accuracy in predicting energies and forces.

\subsection{Atomic modeling of amorphous TiO$_2$-doped GeO$_2$ }

The training set of the machine learning potentials are crystalline GeO$_2$, crystalline TiO$_2$, and amorphous TiO$_2$-doped GeO$_2$ with different doping concentrations. The training structures contain energy-relaxed stable atomic structures as well as structures in which atoms are randomly displaced and unit cell size is expanded and compressed. Molecular dynamic simulation trajectories are also included in the training up to $2000\K$. 

One of the most informative atomic structure characterizations from experiments is the grazing-incidence
pair distribution function (GIPDF), which shows the pair distribution functions in amorphous oxides. The target of modeling is to match the experimental observations as much as possible to give atomic models which can represent the real amorphous materials. In order to generate realistic atomic models, the reverse Monte Carlo (RMC) method has been applied to fit the experimental GIPDF in Ta$_2$O$_5$ and ZrO$_2$-doped Ta$_2$O$_5$ in previous work.\cite{prasai2019high,mishkin2023hidden,jiang2023amorphous} In this work, we generate the models only relying on DFT energy calculations, from classical and machine learning potentials trained based on first-principles calculations. We calculated the X-ray PDF from TiO$_2$-doped GeO$_2$ with 18,000 atoms with different TiO$_2$ doping concentrations (11\%, 27\%, 44\% and 48\%). From the results, we find our models can reproduce an X-ray PDF that agrees well with experimental observation without any RMC fitting.

The metal-oxygen network can be modified by TiO$_2$ doping concentrations. In amorphous TiO$_2$-doped GeO$_2$, a Ge atom prefers to be 4 coordinated by O atoms while Ti prefers a more than 4 O atom coordination, such as 5 or 6. The different local atomic structures of GeO$_2$ and TiO$_2$ compete with each other, resulting in a metal-oxygen first RDF peak center shift. The Ge-O bond shifts to a longer bonding length and Ti-O to shorter bonding compared to pure a-GeO$_2$ and a-TiO$_2$. These findings are consistent with the experimental observations in Raman and RDF measurements\cite{osovsky2023atomic,bhowmick2024unveiling}.

\begin{figure}
	\centering
	\includegraphics[width=0.48\textwidth]{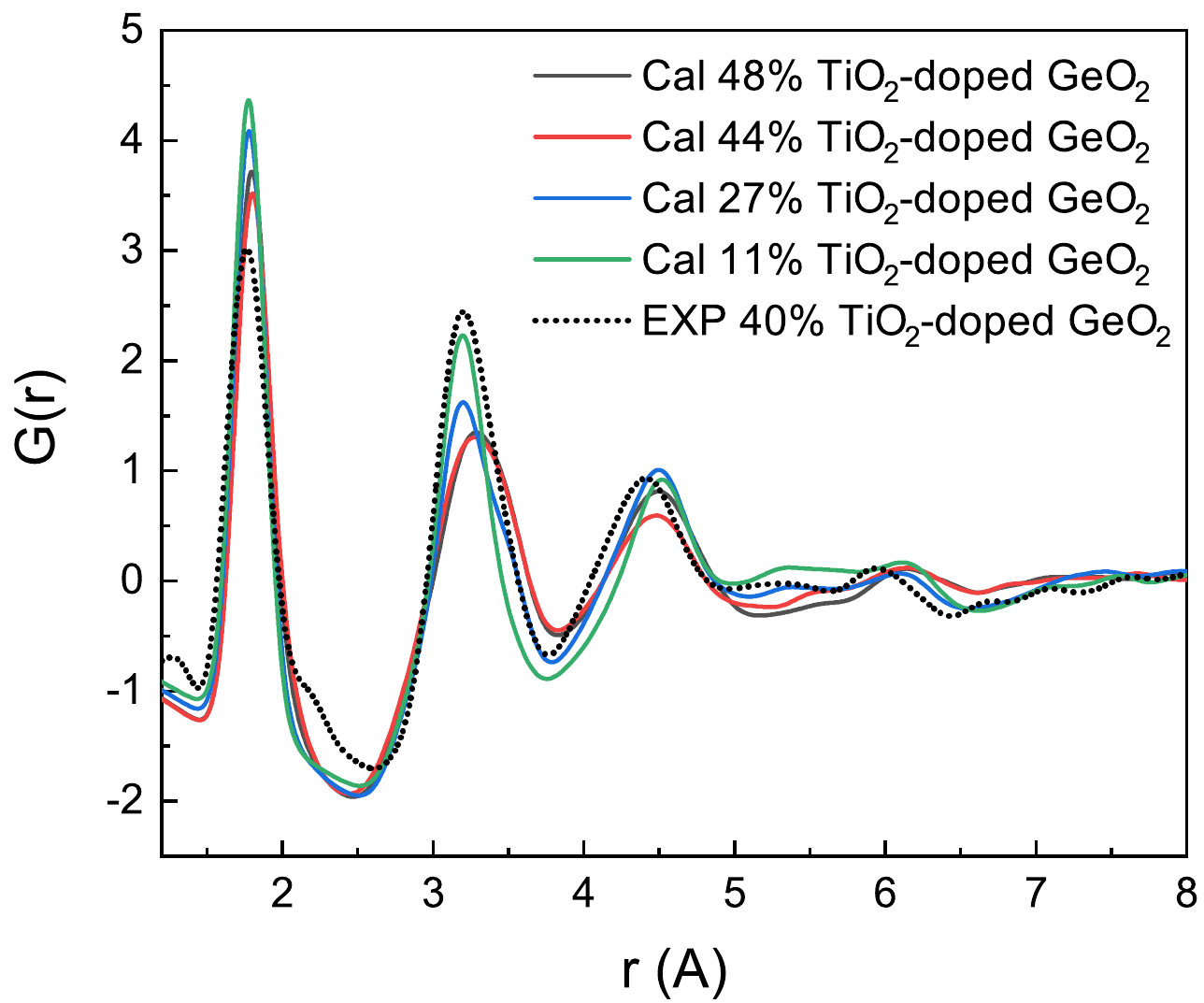}
	\caption{The calculated X-ray RDF (solid lines) of TiO$_2$-doped GeO$_2$ from atomic models with 18,000 atoms and the experimental X-ray RDF (dotted line) of as-deposited TiO$_2$-doped GeO$_2$ for doping levels as identified. The 40\% TiO$_2$-doped GeO$_2$ amorphous sample was deposited using ion beam sputtering by Carmen Menoni's Group at Colorado State University.}
	\label{xrayrdf}
\end{figure}

\subsubsection{Mass density in doped amorphous oxides}

\begin{figure}
	\centering
	\includegraphics[width=0.49\textwidth]{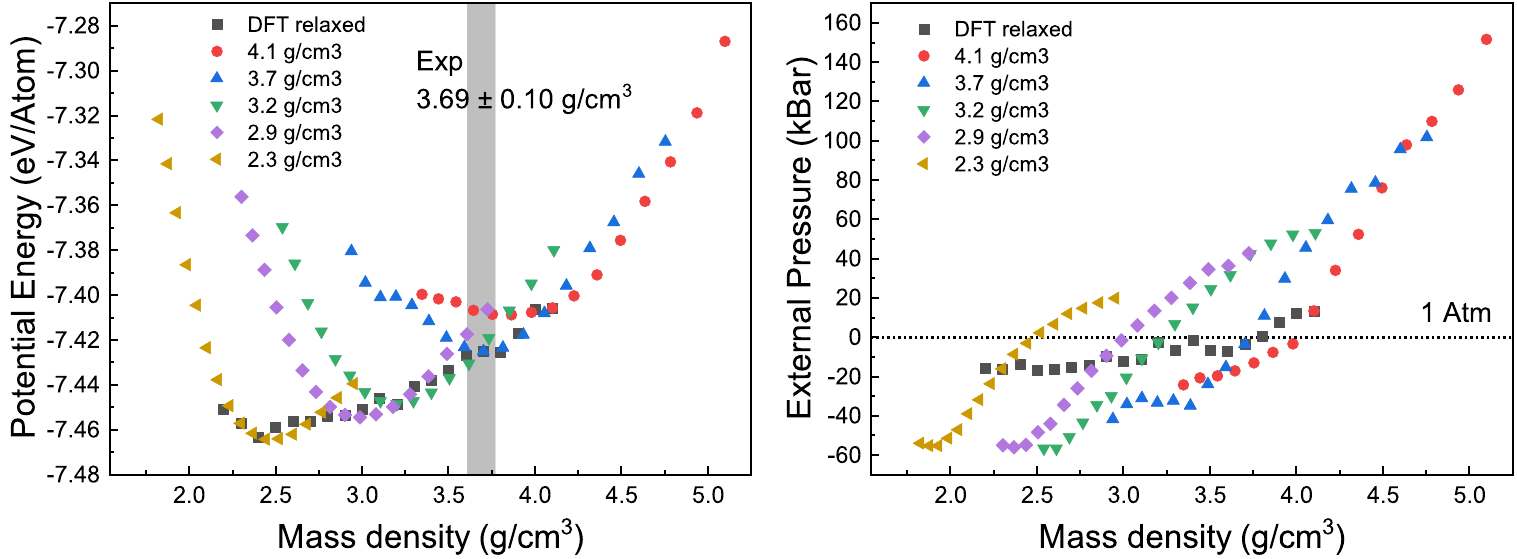}
	\caption{Potential energy and external pressure vs. mass density from DFT calculations. The grey region shows the experimental density of 44\% TiO$_2$-doped GeO$_2$.}
	\label{pot}
\end{figure}

Different from a crystal, an amorphous material has no long-range order and varies in mass density. The atomic structures not only depend on the preparation conditions, but also the thermal history. Unlike crystalline materials, amorphous materials can be stable at different mass densities below the temperature at which no more unstable modes exist\cite{sastry1998signatures,la2000instantaneous,angell2000relaxation}. To better understand this behavior, we generated a series TiO$_2$-doped GeO$_2$ models with different mass densities and minimized their DFT energies to the local minimum. These models are initially randomly generated with fixed volume in equilibrium at high temperature ($2500\K$) using classical potentials and cooled to $1000\K$ within $1\ns$. Then the models are quenched to $0\K$ physical short-range order, which is guaranteed by classical potentials. These atomic structures are then further optimized by minimizing their potential energies using DFT calculations. Here, we use DFT calculations instead of pre-trained MLP to get more accurate atomic structures, since we know MLP can be inaccurate when applied to amorphous models with mass densities different from training. Each structure model has 720 atoms in total. The potential energy vs. mass densities of these models are plotted in Fig.~\ref{pot}. We also performed uniform volume expansion and compression followed by DFT energy minimization starting from these models. The potential energy response to the mass density changes is extracted. 

From the results in Fig.~\ref{pot} we can see that all the relaxed models are indeed at local minima, indicating that these structures are stable for each mass density while the underlying potential energy continually decreases from $4.1\gcm^3$ to $2.4\gcm^3$. We also find that large volume change can cause the atomic structure to reorganize inelasticially such as expanding models from $3.7 \gcm^3$ to $3.4 \gcm^3$ (Fig.~\ref{pot} blue triangles), where the potential energies introduce an additional flat region. To predict the most probable experimental mass density from these calculations, we search for the most symmetric potential energy well, where the transition probabilities to the higher density and lower density structures are likely to be equal, assuming the number of possible atomic configuration transitions are constant near each local minimum. For 44\%-TiO$_2$-doped GeO$_2$, we find that the most symmetric potential energy wells are near $3.7 \gcm^3$, which is consistent with the experimental measured density\cite{vajente2021low}. With this simple assertion, we can obtain the most probable experimental mass density from the calculations.

We also find that the amorphous models with different densities can have different residual strains. These residual strains can be reduced by changing the volume and further relaxing atomic structure. More detailed atomic structure analysis shows that local strains can still exist even when the strains are zero at the boundaries, which indicates that the amorphous oxides have a more complex local atomic structure, as discussed in the following section.

\subsubsection{Void analysis}

\begin{figure}
	\centering
	\includegraphics[width=0.49\textwidth]{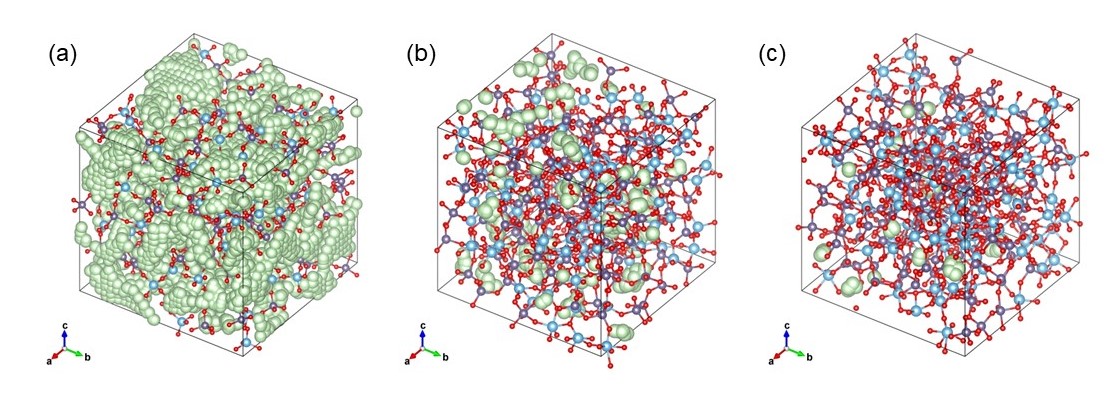}
	\includegraphics[width=0.49\textwidth]{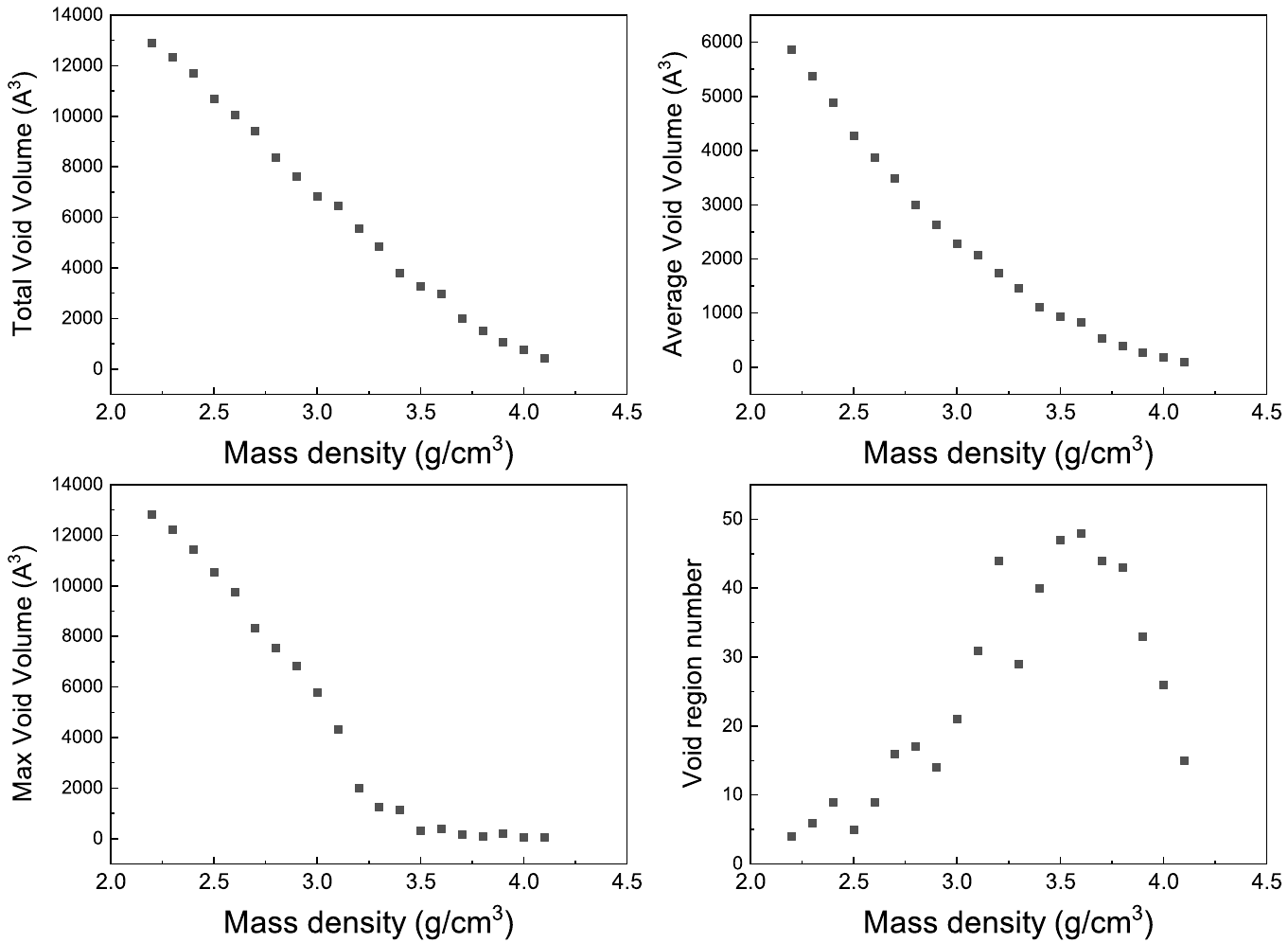}
	\caption{Voids within amorphous 44\% TiO$_2$-doped GeO$_2$ models with different mass densities. The top row shows structures at (a) $2.3 \gcm^2$, (b) $3.7 \gcm^2$, (c) $4.1 \gcm^2$; spheres show atoms or voids (red-O, blue-Ti, purple-Ge and green-voids).  Plots show total, average, and max void volumes and void number, as function of mass density.}
	\label{voids}
\end{figure}

The mass density variations in amorphous materials are related to the different distributions of the empty spaces (or voids) within the metal-oxygen networks, lower densities corresponding to larger unoccupied space within the material. Our void analysis identifies void regions and their size distribution within models. All atoms are considered to be spheres with an empirical covalent radius of \cite{slater1964atomic} $1.25\tAA$ for Ge; $1.40\tAA$ for Ti; and $0.60\tAA$ for O. An efficient algorithm is developed to find empty spaces where we can insert a foreign sphere with a given radius; $1.45 \tAA$ is chosen to find spaces where an Ar atom can fit. Three examples are shown in Fig.~\ref{voids}, where Ge atoms are shown in purple, Ti atoms in blue, O atoms in red, and voids are characterized by the light green surface of a series of spheres.

In atomic models of amorphous oxides, the total void space volume is correlated with the mass density; the lower the mass density is the larger the total void volume is. At high densities, the voids are relative small in size. As mass density decreases the number of voids at first increases with void size not changing significantly. As the mass density further reduces, the size of voids starts to increase and small voids combine into larger voids, resulting in a reduction of total void number (see Fig.~\ref{voids}). From void analysis, we find that the number of unconnected empty space regions peaks when the mass density approaches the experimental density in 44\% TiO$_2$-doped GeO$_2$.

\begin{figure*}
	\centering
	\includegraphics[width=0.85\textwidth]{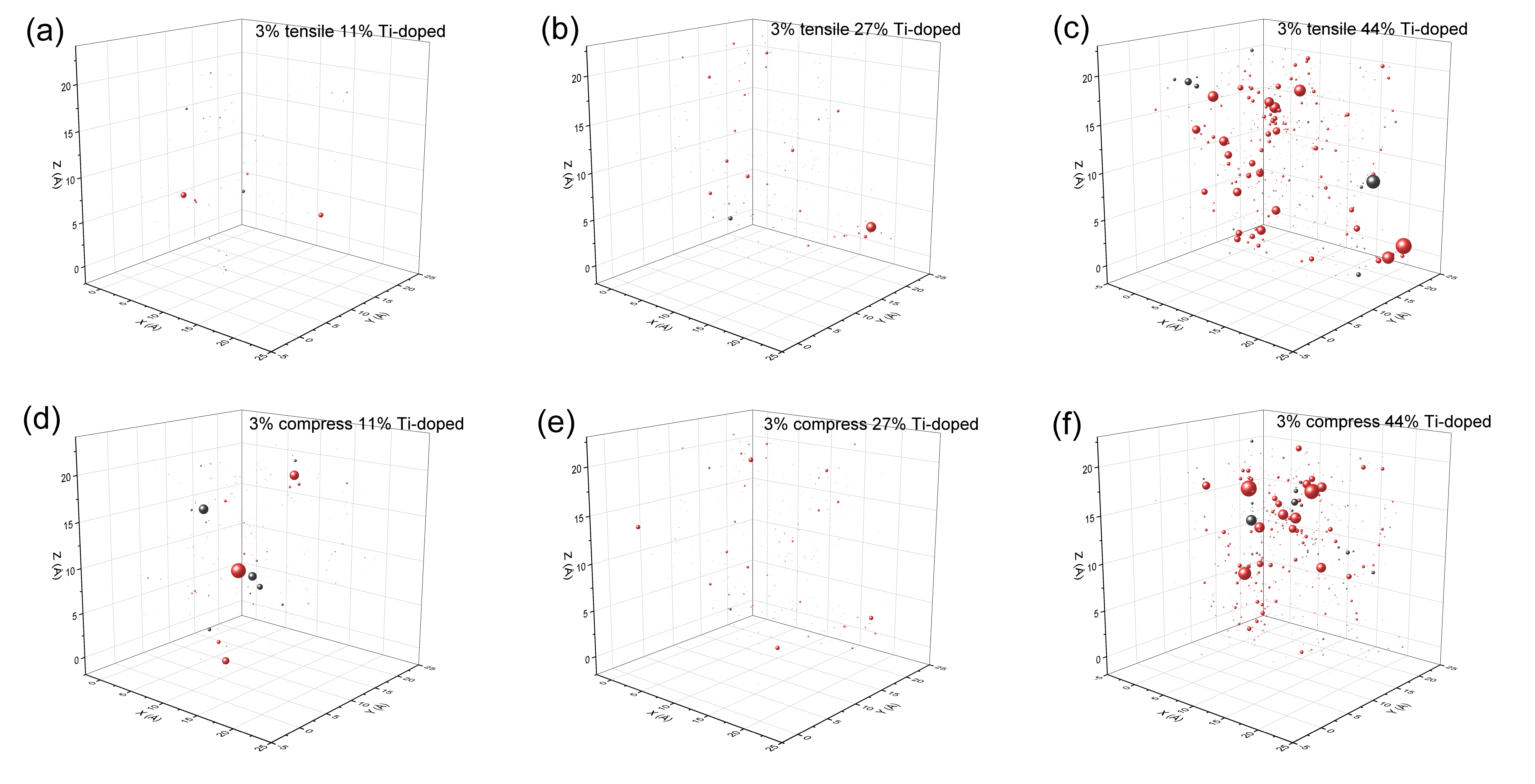}
	\caption{Black spheres represent Ge-O bonds and red spheres are Ti-O bonds. The radii are scaled based on the bond length change as a percentage from initially non-strained to strained configurations. (a), (b), and (c) are 3\% tensile strained; and (d), (e), and (f) are 3\% compressive strained for 11\%, 27\% and 44\% TiO$_2$-doped GeO$_2$ respectively.}
	\label{bondchange}
\end{figure*}

\subsubsection{Additional Ar and O in the amorphous oxides}

In the previous section, void regions have been identified where foreign elements can be inserted according to their classical atomic radius. In experiments, Ar atoms are often trapped in the film during the ion beam deposition process\cite{paolone2021effects,abernathy2021exploration,paolone2022argon}. From previous study, about 3~at.\% Ar atoms were found in ZrO$_2$-doped Ta$_2$O$_5$\cite{abernathy2021exploration} and 1~at.\% in TiO$_2$-doped GeO$_2$\cite{lalande2024ar}. The percentage of argon can be reduced by thermal annealing. These additional Ar in the metal-oxygen network are likely filling up or trapped in these voids. The oxygen concentration can also deviate from stoichiometry as-deposited and change during thermal annealing\cite{abernathy2021exploration}. The voids can potentially accommodate O by reorganizing the metal-oxygen network, which can fix an oxygen deficiency during annealing.

To study the effect of added atoms on the atomic structures, an O or Ar atom is added into one of the possible void regions of fully DFT-optimized 44\% TiO$_2$-doped GeO$_2$ models with a density of $3.7\gcm^3$. The atomic structures with an additional O or Ar are further optimized by minimizing their energies based on DFT calculations. From the optimized models, we find an added O atom in the voids forms bonds to nearby metal (Ge or Ti) atoms. Some of these can create metal-oxygen-oxygen-metal chains where the oxygen-oxygen distance is 1.4--$1.5\tAA$. Such O-O pairs are stable and do not show spin polarization in first-principles calculations with PAW-PBE functionals. The addition of Ar atoms in the voids does not change the metal-oxygen bonds in the networks and only slightly distorts the bond angles around the Ar atoms. The cohesive energy of adding one oxygen atom into 44\% TiO$_2$-doped GeO$_2$ is negative and adding one Ar is always positive, which means thermal annealing with oxygen may change the stoichiometry of the oxides depending on the atmosphere and the trapped Ar atoms tend to escape. These results are consistent with the experimental observations.\cite{abernathy2021exploration,lalande2024ar}

\subsection{Elastic modulus and strains}

\begin{table*}
	\def\arraystretch{1.1}
	\setlength{\tabcolsep}{0.8em}
\caption{Elastic moduli of TiO$_2$ doped GeO$_2$ from EME-SNAP models and experimental measurements.}
	\begin{tabular}{ccccccc} \toprule
		Ti:GeO$_2$  & Ti [\%] & Density ($\gcm^3$) & Young's (GPa)   & Bulk (GPa) & Shear (GPa) & Poisson ratio \\ \hline
		EXP\cite{vajente2021low}    & 44.6 $\pm$ 0.3 & 3.69 $\pm$ 0.10  & 91.5 $\pm$ 1.8 & -- & -- &  0.25 $\pm$ 0.07  \\
		EME-SNAP    & 44 & 3.7 & 83 & 50 & 34 &  0.23  \\ \hline
	\end{tabular}
	\label{elastic}
\end{table*}

Amorphous oxides can be stable at different mass densities, but they respond to external strains differently. From Fig.~\ref{pot}, we find that the models with low mass densities are more easily compressed compared to stretched, due to the steeper potential energy increase with tensile strain compared to compressive strain; while at high mass densities, stretching is easier than compression. The metal-oxygen networks in the amorphous oxides will go through non-elastic deformation when large tensile strain is applied. For an example, the potential energy of TiO$_2$-doped GeO$_2$ models (initially relaxed at $3.7 \gcm^{3}$) flattens when stretched to $3.4 \gcm^{-3}$.

Bulk modulus, Young's modulus and shear modulus are response functions to deformations of the material along different directions. It is computationally very intensive to calculate elastic properties from first-principles. Machine learning potentials (MLP) that are accurate enough to predict the elastic properties are the best choice for treating amorphous oxides. The elastic moduli are thus calculated from the elastic strain tensors according to the Voigt-Reuss-Hill approximation.\cite{chung1968voigt,ray1984statistical,shinoda2004rapid,clavier2017computation} By changing the simulation box size along different directions, the elastic strain tensors are calculated from the change of the system energies under strain. The calculated elastic moduli with machine learning potentials are shown in Table~\ref{elastic}. Unlike for crystalline materials, the finite difference method can hardly be used to calculate the elastic modulus of amorphous materials because of large numerical noise. The amorphous materials can easily go to another nearby local minimum with small atomic displacements, which leads to an incorrect elastic modulus, sometimes even a negative value. In this study we perform MD simulations at finite temperature for a given value of strain and use the average value of potential energy to estimate the strain tensors. The so-obtained elastic moduli are thus robust. The calculated elastic moduli using machine learning potentials are consistent with the experimental measurements. Nevertheless, this method is not feasible for DFT elastic modulus calculations since DFT calculation costs are a factor more than $10^3$ times greater as compared to SNAP machine learning potentials.

It is known that the amorphous materials are macroscopically uniform, while at atomic level the metal-oxygen network is not locally uniform. The metal-oxygen bonds are distributed over a sizable range, reflecting compressive or tensile strain caused by surrounding atoms or void spaces. Here we show these bonding strain distributions by representing each bond as a sphere. The center of the sphere is the middle point of the metal and the oxygen atom. Assuming the most probable bond length is zero strain, the red color represents compressive strain and the blue tensile strain. The radius of the sphere is proportional to the deviation of the most probable bond length shown in Fig.~\ref{bondchange}.

The results show that there are both compressive and tensile local strained regions in the amorphous metal oxygen networks. These local strains in the models are stable and are not eliminated by simply relaxing the atomic structure with DFT energy minimization.  Local strains still exist after equilibrating at $300\K$ for $10\ns$ in the molecular dynamics (MD) simulations. The ``bumpy'' energy landscape of amorphous materials can stabilize these local strains and protect them for a very long time at room temperatures. These local strains are not necessarily correlated with the external stress, they exist even when the external stress is zero. Thermal annealing can speed up relaxations and release some of the local strains, which can be observed experimentally by measuring the full width at half maximum (FWHM) of the first RDF peak \cite{yilmaz2008pathways,thomas2019atomistic}. The narrower the FWHM is, the fewer strained bonds are in the metal-oxygen network. These local strains are found to be correlated with mechanical losses in TiO$_2$-doped GeO$_2$.\cite{bassiri2013correlations}

With additional strains applied, the non-uniform elastic response can be extracted by tracking the atom-atom distance change. For low density amorphous models, there is little bond distance change and the metal-oxygen networks adapt to the external strains by mainly changing the bond angles; at high densities, the Ti-O bond length changes are the major response to the applied strains. With the low TiO$_2$ doping concentrations, changes are suppressed, and when the TiO$_2$ doping concentration is higher than 27\%, the Ti-O distances are changed significantly due to strains. The metal-oxygen bond change distributions are plotted in Figure~\ref{bondchange}. 

\section{Summary}

Amorphous TiO$_2$-doped GeO$_2$ is a promising coating material for LIGO mirror applications. These coatings will play a crucial role in reducing noise and improving sensitivity in gravitational-wave detectors. However, modeling these doped amorphous oxides is challenging, due to their complex energy landscape. To address this, our study has focused on modeling and understanding atomic structures. By combining the first-principles calculations and machine learning potentials, good atomic models are generated with different doping concentrations and mass densities. We generate classical potentials that contain three-body interactions to improve Ge-Ge correlations compared to existing classical pair potentials to accelerate model generation and assist the model generation especially at high temperatures. Additionally, we also apply the on-the-fly machine learning potentials to refine atomic structures of large models from classical potentials. Balancing accuracy and speed, we further developed EME-SNAP and MTP potentials that allow us to prepare and study amorphous models with 18,000 atoms and to perform MD simulations at nanosecond timescale. Although we have tried to further improve the machine learning potentials used in this work, they are still limited to applications within certain doping concentrations and temperature ranges. With amorphous models with different mass densities, we investigate the pattern of the metal oxide network and the evolution of voids for the doped amorphous oxides materials. The richness and complex networks of doped amorphous oxides gives us greater control over material properties for a wider range of applications. Our work provides an advanced way of modeling and generating insights from atomic structures for complex amorphous materials.

\begin{acknowledgments}
This work is supported by the National Science Foundation (NSF) under Award Nos. PHY-2011776, PHY-2011770, PHY-2309291 and PHY-2429331 associated with the LSC Center for Coatings Research.   Computations were performed using the utilities of the National Energy Research Scientific Computing Center (NERSC) and the University of Florida Research Computing HiPerGator.

\end{acknowledgments}

\bibliography{ref}

\end{document}